\documentstyle[aps,twocolumn,rotate,epsf]{revtex}

\author{Yan V. Fyodorov$^{a,c}$,  Dmitry V. Savin$^{b}$
  and Hans-J\"{u}rgen Sommers $^{a}$}
\address{$^a$ Fachbereich Physik, Universit\"{a}t-GH Essen, Essen
  45117,Germany}
\address{$^b$ Budker Institute of Nuclear Physics,
  630090 Novosibirsk, Russia}
\address{$^c$ Petersburg Nuclear Physics Institute,
  Gatchina 188350, Russia}
\title{Parametric Correlations of Phase Shifts and Statistics of  Time
  Delays in Quantum Chaotic Scattering: Crossover between Unitary
  and Orthogonal Symmetries}
\date{\today}

\begin{document}
\draft
\maketitle

\begin{abstract}
We analyse universal statistical properties of
phase shifts and time delays for open chaotic systems
in the crossover regime of partly broken time-reversal invariance.
In particular, we find that the distribution of the time delay shows $\tau^{-3/2}$
behavior for weakly open systems of any symmetry.

\end{abstract}

\pacs{}

The energy-dependent scattering phase shifts $\theta_a (E)$
defined via
 the eigenvalues $\exp{i\theta_a}\,;\,\, a=1,...,M$ of the
$M\times M$ unitary scattering matrix $\hat{S}(E)$ are important
and frequently used characteristics of the process of quantum
scattering.
In particular, the derivatives of phase shifts over 
energy
 $\tau_a=\partial \theta_a/\partial E$ are related to the
duration of a collision
 event. For example, the quantity
$\tau_W=M^{-1}\sum_a\tau_a$ is the typical time delay due to
scattering, the so-called Wigner-Smith time delay
\cite{WignerSmith}. When some external parameters are taken into
consideration (e.g. a magnetic field) the corresponding
parametric
variation of the phase shifts can as well be related 
to some observables \cite{Akk}.

Growing interest to the
universal features \cite{AlSi}
of quantum systems whose classical
counterparts demonstrate chaotic dynamics attracted considerable
attention to the
process of quantum chaotic scattering, see
\cite{chaosc,Gasp}
and references therein. From this point of view,
different statistical characteristics of phase shifts and time delays
were addressed in experiments on chaotic microwave
reflection\cite{Dosmifr} as well as in several numerical studies of
various models of quantum scattering in disordered and chaotic
systems\cite{Shushin,JPich,Dietz,Wang}.

It is interesting to mention that for
the case of only two open channels $M=2$ the phase
shifts $\theta_{1,2}$
can simply be related to the phases of
transmission and reflection coefficients, see e.g.\cite{antiThoul}.
The latter quantities are amenable to direct
experimental
measurements in quantum dots, see \cite{dotphase} and references
therein.  Another
fact attributing additional interest to studies of time delay
statistics is that it is
intimately connected with the issue of
mesoscopic fluctuations of
dynamic admittances of microstructures\cite{Gopar}.

One can extract statistical characteristics of
the S-matrix exploiting a semiclassical periodic
orbit expansion like that
provided by the Gutzwiller trace formula, see
examples of such calculations
in \cite{chaosc,Gasp,Dosmifr}. To
proceed in this way, one has to employ some approximations (most
frequently, the so called diagonal approximation).
The resulting
expressions provide an important insight into the problem.
In particular, the semiclassical approximation for the time delay
correlations at two different energies was derived by
Eckhardt\cite{Eckhardt}.
However, the results obtained in such a way
have a restricted domain of applicability; in particular, they
fail
to describe the system with only few open channels: $M\sim 1$.

A
powerful alternative to the semiclassical methods in extracting the
{\it universal} (i.e. generic and system independent )
statistical characteristics of the scattering matrix is provided by
the random matrix approach. In particular,
in the "Heidelberg variant" of this approach \cite{VWZ} one relates
the scattering matrix $S(E)$ to the Hamiltonian of a closed
counterpart of the
open system. The latter Hamiltonian is considered
to be a member
of an ensemble of random matrices of appropriate
global symmetry - an idea
commonly accepted in the domain of
Quantum Chaos. The use of random matrices gives
one the possibility to apply the very powerful machinery of
averaging developed by Efetov\cite{Efrev} and to calculate different
correlation functions explicitly for any number of open channels and
arbitrary coupling to continua.

When employing the Heidelberg method the actual
calculation depends quite essentially on the symmetry of the
underlying Hamiltonian. The simplest case to study
 corresponds to completely broken time reversal invariance (TRI)
(systems in strong enough magnetic field), when the random
matrix Hamiltonian $ \hat{H}$ is taken from the
Gaussian Unitary Ensemble (GUE). For such systems
the statistics of phase shifts, delay times and resonance poles
was thoroughly investigated by two of us recently\cite{FS}, see
a detailed exposition of the calculation in\cite{FSrev}. In the
opposite case of fully preserved TRI when
$\hat{H}$ is a member of the Gaussian Orthogonal Ensemble (GOE)
some aspects of time evolution of a chaotic system were considered in
\cite{DHM-92},
the correlation function of Wigner-Smith time delays for
two different energies was
found in \cite{LSSS} and the distribution of time delays was
obtained for the perfect coupling case in \cite{Gopar},
see also\cite{SZZ}. Let us also mention the paper\cite{Macedo}
addressing the
issue of parametric correlations for $S-$ matrix elements.

In the present communication we extend the analysis of
 statistical properties of
phase shifts and time delays to the whole crossover region
of gradual breaking of the TRI.
Different characteristics of chaotic and disordered systems
in this crossover regime
were under quite an intensive theoretical investigation
recently\cite{crth1,creigf,sus}.

Within the framework of random matrix theory, Hamiltonians of the
{\it closed} chaotic systems under consideration are conveniently
represented as\cite{crth1,sus}:
$\hat{H}(y)=\hat{H}_S+i\frac{y}{\sqrt{N}}\hat{H}_A$,
where $\hat{H}_S$ is $N\times N$ GOE matrix and $\hat{H}_A$ is
 a real random antisymmetric matrix of the same dimension.
For the sake of generality
the symmetric matrix $\hat{H}_S$ is taken in the form\cite{AlSi}:
$\hat{H}_S=\hat{H}^{(0)}_S+\frac{x}{\sqrt{N}}\hat{H}^{(1)}_S$.
This form allows one to simulate the influence of such perturbations
(e.g. a variation of the strength of scattering potential)
which do not break the TRI. All elements of random matrices 
are independent and normalised in
such a way that $\left\langle\mbox{Tr}\left( \hat{H}^{(0,1)}_{S,A}\right)^2
\right\rangle=N$.

In the limit $N\to \infty$
the crossover is driven
by the parameter $y\in[0,\infty)$, with $y=\infty$ corresponding to
completely broken TRI.
Physically the
parameter $y$ is proportional to the magnetix flux
through the system $\Phi$.
 One may also notice that the typical shift of the levels
due to the antisymmetric perturbation is
$\delta E_{y}/\Delta\sim y^2$\cite{sus}, where $\Delta$ is the mean
level spacing.

  Within the framework of the Heidelberg
approach \cite{VWZ}
the coupling of the chaotic region to
the incoming/outgoing waves is described with the help of
 the $M\times N$
matrix $\hat{W}$ of amplitudes $W_{ia},\,\, a=1,2,...,M;\quad i=1,...,N$,
which couple the internal
motion to $M$ open channels. In what follows we consider the case
of arbitrary, but fixed $M$ whereas $N\to \infty$.
Without much loss of generality these
amplitudes can be chosen in a way ensuring that the
average
$S-$matrix is diagonal in the channel basis:
$\left\langle S_{ab}\right\rangle=\delta_{ab}\langle S_{aa}\rangle$. The strength of
coupling to continua is convenient to be characterized via the
"sticking probabilities"
(also called the "transmission coefficients")
$T_{a}=1-|\langle S_{aa}\rangle|^2$ which
are given for the present model by the following
expression\cite{VWZ}:
\begin{equation}\label{trans}
 T_{a}^{-1}=\frac{1}{2}
  \left[1+\frac{\gamma_a+\gamma_a^{-1}}{2\pi\nu(E)}
  \right] \, ; \quad
  \gamma_a=\pi\sum_{i}W^{*}_{ia}W_{ia}
\end{equation}
with $\nu(E)=\pi^{-1}(1-E^2/4)^{1/2}$ being the density of states
for the GOE matrices related to the local mean level spacing as
$\Delta=(\nu N)^{-1}$.
The quantity $T_{a}$ measures the part of the flux in channel $a$ that
spends a substantial part of time in the interaction
region\cite{VWZ}.
We see that both limits
$\gamma_a\to 0$ and
$\gamma_a\to \infty$ equally correspond to the weak effective coupling
regime $T_{a}\ll 1$ whereas the
strongest coupling (at fixed energy
$E$ ) corresponds to the value $\gamma_a=1$. The maximal possible
coupling corresponding to the upper bound $T_a=1$ is achieved in the
present model for an energy interval in the vicinity of the center
$E=0$. Below we restrict our attention to this point of spectrum in
order to present our final results in the most compact form.
 Moreover, we consider all channels to be
statistically equivalent: $\gamma_a=\gamma$ for $a=1,...,M$.
Generalization to arbitrary $E$ and non-equivalent channels
can easily be done, see \cite{FSrev}.

In the earlier work \cite{FS,FSrev} it was shown that
one can study very effectively the
statistics of phase shifts $\theta_a$
considered modulo $2\pi$.
To this end we find it to be convenient to introduce the auxiliary
"phases" $\phi$ related to the phase shifts $\theta$ as
$\phi=\arctan{\{\gamma^{-1}\tan{(\theta/2)}\}}$ and consider the density
$\rho_{E,x,y}(\phi)=M^{-1}\sum_a\delta\left(\phi-\phi_a(E,x,y)\right)$.

 The connected part of the correlation function of these densities
is our main object of interest. It can be found performing calculations
{\it mutatis mutandis}, similar to that presented
in\cite{FS,FSrev} and it turns out to be
dependent only on the difference $\phi=\phi_1-\phi_2$:
\begin{eqnarray}\label{phi}
&& K^{\phi}_{\omega,x,y_1,y_2}(\phi) =
 \langle\rho_{E=0,x=0,y_1}(\phi_1)
 \rho_{E=\Omega,x,y_2}(\phi_2)\rangle_c = \nonumber \\
&& \mbox{Re} \int\limits_{-1}^1\!d\lambda\!
 \int\limits_1^{\infty}\!d\lambda_1 \!
 \int\limits_1^{\infty}\!\frac{d\lambda_2}
 {{\cal R}^2} {\cal F}_M(\phi)
 e^{-\frac{x^2}{2}(2\lambda_1^2\lambda_2^2-\lambda_1^2-
 \lambda_2^2-\lambda^2+1)}\times \nonumber \\
&& \mbox{\ }
 e^{-i\omega(\lambda_1\lambda_2-\lambda)+y_1y_2(\lambda_1^2-
 \lambda_2^2) -
 \frac {1}{2}(y_1^2+y_2^2)(\lambda_1^2+\lambda_2^2-\lambda^2-1)}
 \times \nonumber \\
&& \mbox{\ }
   \left\{ (1-\lambda^2)\cosh{\alpha} -
 (\lambda_1^2-\lambda_2^2)\sinh{\alpha} + \right. \nonumber \\
&& \mbox{\quad}
 {\cal R}\left[ (y_1^2+y_2^2) \left( (1-\lambda^2)\cosh{\alpha}+
 (\lambda_2^2-\lambda_1^2)\sinh{\alpha} \right) +
  \nonumber  \right. \\
&&  \mbox{\quad} \left. \left.
 2y_1y_2(\lambda_1^2+\lambda_2^2+\lambda^2-1)\sinh{\alpha}
  \right]  \right\}
\end{eqnarray}
 where the "channel factor" is equal to
\begin{equation}\label{ch}
{\cal F}_M(\phi)=-\frac{\partial^2}{\partial\phi^2}
\left[\frac{(1+i\lambda\tan{\phi})^2}
{1+2i\lambda_1\lambda_2\tan{\phi}
-\tan^2{\phi}(\lambda_1^2+\lambda_2^2-1)}\right]^{M/2}
\end{equation}
and where ${\cal R},\alpha,\omega$ denote
${\cal R}= \lambda_1^2+\lambda_2^2+\lambda^2
-2\lambda\lambda_1\lambda_2-1,\quad\alpha=y_1y_2(1-\lambda^2)$
and $\omega=\pi\Omega/\Delta$.

The correlation function presented above is a very  informative
object. First of all, having it at our disposal it is a
relatively easy task to show
\cite{FSrev} that the correlation function of Wigner-Smith time delays:
$K^{\tau}_{\omega,x,y_1,y_2}=\langle \tau_W(E=0,x=0,y_1)
 \tau_W(E+\Omega,x,y_2)\rangle/\langle \tau_W (E)\rangle^2$
is given in the crossover regime by the same expression Eq.(\ref{phi})
provided one
replaces the "channel factor" Eq.(\ref{ch}) by:
\begin{equation}\label{tau}
 {\cal F}_M^{\tau}=
(\lambda_1\lambda_2-\lambda)^2\left[\frac{(g+\lambda)^2}{
(g+\lambda_1\lambda_2)^2-(\lambda_1^2-1)(\lambda_2^2-1)}\right]^{M/2}
\ ,
\end{equation}
here $g=2/T-1$, with $T$ being the transmission
coefficient introduced above.

Secondly, one can extract explicitly
the general distribution function of the scaled partial
delay times ${\cal P}(\tau_s)=\left\langle\frac{1}{M}\sum_a\delta
\left(\tau_s-\frac{\Delta}{2\pi}\tau_a(y)\right)\right\rangle$
in the crossover regime:
\begin{equation}\label{dist}
\begin{array}{l}
{\cal P}(\tau_s)=
\frac{C_M}{\tau_s^{(M+5)/2}}\int_{-1}^1d\lambda\int_1^{\infty}d\lambda_2
\times \\
\lambda_2^{\frac{M+3}{2}}(\lambda_2^2-1)^{\frac{1-M}{4}}e^{-2y^2
(\lambda_2^2-1)}{\cal J}_1(\lambda_2){\cal J}_2(\lambda,\lambda_2)
\end{array}\end{equation}
where $C_M=\left[(2\pi)^{1/2}2^{M/2+1}\Gamma
\left(\frac{M}{2}+1\right)\right]^{-1}$ and
$$
\begin{array}{l}
{\cal J}_1(\lambda_2)=\int_{0}^{\pi}d\psi
v(\psi)^{\frac{M+1}{2}}
e^{-\frac{\lambda_2^2}{\tau_s}v(\psi)}I_{\frac{M-1}{2}}
\left[\frac{\lambda_2\sqrt{\lambda_2^2-1}}{\tau_s}v(\psi)\right]\\
{\cal
J}_2(\lambda,\lambda_2)=4y^2\left[(1-\lambda^2)e^{-\beta}+\lambda_2^2
(1-e^{-\beta})\right]-\left(1-e^{-\beta}\right)

\end{array}
$$
where $v(\psi)=g-\sqrt{g^2-1}\cos{\psi}$, $\beta=2y^2(1-\lambda^2)$
and $I_p(z)$ stands for the modified Bessel function.

The distribution Eq.(\ref{dist}) is valid for any number of
open channels $M$ and any value of
transmission coefficient $T$ and as such is quite
complicated. To get a better understanding of its typical features
it is reasonable
to look separately at two limiting cases of strong/weak coupling to
continua.

For the strong  coupling regime $T=1$ (i.e. $g=1$) the $\psi-$
integration in
Eq.(\ref{dist}) drops out, but the resulting expression is still
quite cumbersome. However, one can easily find
 the long time asymptotics to be of the following form:
\begin{equation}\label{lt1}
{\cal P}(\tau_s\gg 1)=\left\{\begin{array}{cc}
U_M(y)\tau_s^{-(2+M)}& y>0\\
(2\pi)^{1/2}C_M e^{-\frac{1}{2\tau_s}}\tau_s^{-\left(2+M/2\right)} &
y=0\end{array}\right.
\end{equation}
where $U_M(y)$ is a rather complicated function of the symmetry
breaking parameter $y$.

The second of this expressions holding for
unbroken TRI is actually exact for arbitrary
$\tau_s$ as can be seen performing the limit $y\to 0$ in
 the general Eq.(\ref{dist}) at $T=1$.
This fact was first conjectured in
\cite{SZZ} and derived for the particular case $M=1$
by another method in \cite{Gopar}.

Eq.(\ref{lt1}) demonstrates that the limits $y\to 0$ and
$\tau\to\infty$ do not commute. To understand this phenomenon better
it is instructive to consider the case of "weakly broken" TRI,
$y\ll 1$, in more detail.
A close inspection shows that for such regime
there emerges one more relevant time-scale $\tau_y\propto y^{-2}\gg 1$
such that for the domain $1\ll \tau_s\ll \tau_y$ the distribution
function ${\cal P}(\tau_s)$ shows the GOE-like behaviour:
${\cal P}(\tau_s)\propto\tau_s^{-(M/2+2)}$, whereas at  $\tau_s\gg
\tau_y$
this behaviour
 changes to the GUE-like: ${\cal P}(\tau_s)\propto \tau_y^{M/2}
\tau_s^{-(M+2)}$.

We suggest the following transparent physical interpretation of the
scale $\tau_y$: this is just the time $\hbar/\delta E_y$ necessary by
the Heisenberg uncertainty relation to resolve a typical shift $\delta
E_y$ due to the TRI-breaking perturbation. If the particle dwells
in the scattering domain for a time shorter than $\tau_y$ it can not
"feel" the magnetic field effects and
the corresponding asymptotics is GOE-like. However, for large enough
times the particle resolves the effect of broken TRI whatever small
is the magnetic field. This explains why the most distant asymptotics
of the time delay distribution is always GUE-like, provided the
magnetic field is not identically zero.

Let us now turn our attention to the opposite limit of an almost "closed"
chaotic system: $T\ll 1$. Exploiting  $g\gg 1$ we find for
 arbitrary number of open channels and arbitrary $y$
 the following universal (up to a coefficient)
behaviour of delay time distribution:
\begin{equation}\label{threehalf}
  {\cal P}(\tau_s)\propto g^{-1/2}\tau_s^{-3/2}
  \quad \mbox{when}\quad  g^{-1}\ll \tau_s\ll g
\end{equation}
in the parametrically large region of delay times.
The proportionality  coefficient in this formula depends
on the symmetry breaking parameter $y$ and $M$ in a complicated way.

Such a $\tau^{-3/2}$ behaviour holding irrespective of the TRI symmetry
is the most robust feature of the time delay statistics of 
weakly open
chaotic systems. It was first obtained in \cite{FSrev} for the case
of broken TRS, but physical arguments show that it is a very
general feature simply
following from the picture of well-isolated resonances typical for
such systems
\cite{FSrev}. It also seems to be quite insensitive to the
particular details  of definitions of time delays and holds for
distributions
of such slightly different quantities as Wigner-Smith time delay,
partial delay times or even "dwell times". In Fig.1 we plotted a
typical fluctuating pattern of energy-dependent "dwell times"
obtained in the paper
\cite{Wang} in the course of numerical simulations of quantum
chaotic scattering on a two dimensional cavity in tunneling contact
with two waveguides. Sampling the distribution of dwell
 times over the chosen range of energies we find a good
agreement  with the predicted $\tau^{-3/2}$ behaviour, see Fig.1.

\begin{figure}
\unitlength 1cm
\begin{picture}(8,6)
\epsfxsize 9cm
\put(-1,-2){\rotate[r]{\epsfbox{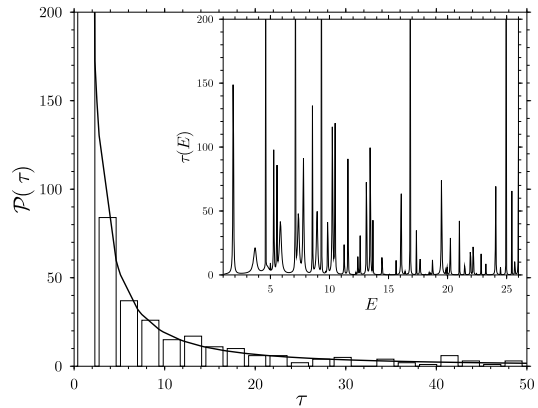}}}
\end{picture}
\unitlength 1bp
\caption{The distribution of the dwell times $\tau$ in chaotic
scattering in
 weakly open Sinai-like billiard. Inset shows a fluctuating pattern
 of dwell time versus energy (data were
kindly provided to us by the authors of Ref.[10]). Solid line shows
the theoretical prediction
${\cal P}(\tau)\propto \tau^{-3/2}$.}
\label{fig1}
\end{figure}

Outside the parametrically large interval $g^{-1}\ll \tau_s\ll g$
our general
expression Eq.(\ref{dist}) predicts an exponential cutoff at $\tau_s
\lesssim g^{-1}$  and a crossover to the behaviour described by
Eq.(\ref{lt1})
for the asymptotically large times $\tau\gg g$. One can check that
for "weakly broken" TRI again there emerges a scale $\tau_y(g)\sim g/y^2$
 such that the asymptotic tail is GOE-like at $g\ll \tau_s\ll
\tau_y(g)$, but
always GUE like for $\tau_s\gg\tau_y(g)$,
in full agreement with the discussion above.

Finally, having in mind the comparison with
the semiclassics let us consider in more detail
the large-channel limit $M\gg 1$ of our general
expressions like Eq.(\ref{phi}) describing the correlations
of phase shifts and Wigner-Smith time delays.
When doing this it is natural to
consider the angle difference $\phi=\phi_1-\phi_2$
to be of the order of $\phi\sim 1/M\ll 1$. Then one
rescales $\phi\equiv
\tilde{\phi}/M$,
substitutes $\tan{\phi}\sim \tilde{\phi}/M$ in the "channel factor"
Eq.(\ref{ch}) and
performs the limit $M\to \infty$ explicitly.
For $\omega=0$
the resulting expression turns out to be {\it identical} to the
parametric correlation function of eigenvalues of large random
matrices in the crossover regime derived for the
first time by  N.Taniguchi et al.  \cite{sus}. Taking into account that
$\phi_1-\phi_2\sim 1/M$ results also in $\theta_1-\theta_2\sim 1/M$,
 we conclude that the statistics of scattering phase shifts in the large
$M$ limit
is just the same as that of energy levels of {\it closed} chaotic
systems.
The latter conclusion is in agreement with the available numerical
results
obtained for  a realistic model of chaotic systems with $M=23$ in
\cite{Dietz}.  It is also interesting, that the only modification required for $\omega\ne 0$ is to replace $\tilde{\phi}\to\tilde{\phi}+\omega$.

Considering the time delay characteristics one should take into account
that  the width of the time delay distribution is of
lower order in $M$ as compared with the mean value $\langle\tau_W\rangle$
 when resonances are overlapping: $MT\gg 1$\cite{LSSS}.
To extract the time delay correlations in
 the corresponding limit requires a calculation similar to that
done in the paper by Pluhar et al. \cite{sus}. The resulting expression
turns out to be quite a transparent one and is given by:
\begin{equation}\label{as}
 K^{\tau}_{\omega,x,y_1,y_2} =
 \frac{1}{2}\left(\frac{\Gamma_{-}^2-\omega^2}
 {\left[\Gamma_{-}^2+\omega^2\right]^2}
 +\frac{\Gamma_{+}^2-\omega^2}{\left[\Gamma_{+}^2+\omega^2
 \right]^2}\right)
\end{equation} provided that
$\Gamma_{\pm}\equiv MT/2+x^2+(y_1\pm y_2)^2 \gg 1$.
Actually, this formula is
nothing else but the semiclassical expression for the time delay
correlator. It can be obtained from the Gutzwiller trace formula in
diagonal approximation, with the quantity $MT/2$ being replaced by the decay rate out
of the chaotic region, see \cite{Eckhardt,FSrev}.

The last point to be mentioned is related to the
issue of  fluctuations of low-frequency admittance as defined
in \cite{Gopar}. We noted above that
the time delay fluctuates weakly in the many-channel limit $M\to\infty$.
Using this fact and the relation between the time delay and
the low-frequency admittance $G^I(\omega)$
presented in \cite{Gopar} (see, however,\cite{note}) one
 finds that the parametric correlator
$\langle G^I(0,y_1)G^I(x,y_2)\rangle/ \langle G^I\rangle^2-1$
of the admittance in the
limit $M\gg 1$ is given by:
\begin{equation}\label{admit}
 \frac{1}{2}\left(\frac{1}{\Gamma_{-}^2}
 + \frac{1}{\Gamma_{+}^2}\right)
 \left(1+\frac{Me^2}{C_e\Delta}\langle\tau_W\rangle\right)^{-2}
\end{equation}
where $C_e$ denotes the so called "geometric capacitance"\cite{Gopar}
and $e$ stands for the electron charge.
In the limiting cases of unbroken ($y_1=y_2=0$) and completely
broken ($y_1=y_2\to\infty$) TRI this expression coincides with that
 found recently by another method
by Brouwer and B\"{u}ttiker \cite{Gopar}.

\narrowtext

We are very greatful to V.V. Sokolov for instructive discussions and to
Prof. Hong Guo for kindly providing us with the
numerical data used to extract the time delay distribution depicted
in Fig.1.
The financial support from SFB 237
der Deutschen Forschungsgemeinschaft as well as from the
program "Quantum Chaos" (grant No. INTAS-94-2058) is acknowledged
with thanks.

\vspace{-1.0cm}.

\end{document}